\documentclass[twocolumn,runningheads,final]{svjour2}
\smartqed  
\usepackage{graphicx}

\newcommand{\beq}{\begin{equation}}
\newcommand{\eeq}{\end{equation}}
\newcommand{\beqn}{\begin{eqnarray}}
\newcommand{\eeqn}{\end{eqnarray}}
\newcommand{\lppr}{\stackrel{<}{\scriptstyle \sim}}
\newcommand{\gppr}{\stackrel{>}{\scriptstyle \sim}}

\journalname{Astrophysics and Space Science}
\begin{document}

\title{Supermassive binary black holes among cosmic gamma-ray sources}

\titlerunning{Supermassive binary black holes}

\author{Frank M. Rieger}  

\authorrunning{Frank M. Rieger} 

\institute{F.M. Rieger \at
           UCD School of Mathematical Sciences \\
           University College Dublin, Belfield, 
           Dublin 4, Ireland\\
              \email{frank.rieger@ucd.ie}}           
       
\date{Received: date / Accepted: date}

\maketitle

\begin{abstract}
Supermassive binary black holes (SBBHs) are a natural outcome 
of galaxy mergers. Here we show that low-frequency ($f \leq 
10^{-6}$ Hz) quasi-periodic variability observed from cosmic 
blazar sources can provide substantial inductive support for 
the presence of close ($d \lppr 0.1$ pc) SBBHs at their centers. 
It is argued on physical grounds that such close binary systems 
are likely to give rise to different (although not independent) 
periodicities in the radio, optical and X-ray/TeV regime, and, 
hence that detection of appropriate period ratios significantly 
corroborates the SBBH interpretation. This is illustrated for
a binary model where optical longterm periodicity is related to 
accretion disk interactions, radio periodicity to Newtonian jet 
precession, and periodicities in the high energy bands to the 
orbital motion of the jet. We use the observed periodicities to
constrain the properties for a sample of SBBH candidates including 
OJ~287 and AO 0235+16, and discuss the results within the context 
of jet activity and binary evolution.
\end{abstract}

\section{Introduction}
Violent galaxy mergers are known to play a vital role in the
cosmological evolution of galaxies and the growth of supermassive 
black holes (BHs). Since almost every bright galaxy seems to 
contain a black hole (BH), frequent formation of supermassive 
binary black holes is naturally expected to occur during cosmic 
time, e.g., see \cite{beg80,fer05,lob05}. In cosmic bottom-up 
scenarios, for example, elliptical galaxies, such as the host 
galaxies of radioloud AGNs, are usually thought to form when two 
spiral galaxies collide and merge. Interacting galaxies are indeed 
observationally well-known and the observation that the relative 
number of spiral to elliptical galaxies tends to increase in 
distant clusters \cite{dre97}, gives additional credit to the 
underlying evolutionary picture. Moreover, direct observational 
evidence for formation of wide (separation $d \gg 1$ pc) 
supermassive binary systems has been established recently based 
on Chandra observations of the ultraluminous infrared galaxy 
NGC 6240 \cite{kom03} and VLBA observations of the radio galaxy 
0402+379 \cite{rod06}. 
While the existence of wide SBBHs may thus be regarded as 
observationally well grounded, the anticipated existence of close 
($d \sim 0.01-1$ pc) SBBHs appears much more ambiguous. Indeed, 
there is little direct observational evidence so far for a close, 
secondary supermassive BH in the nuclear region of Active Galaxies. 
On the other hand, dynamical friction and slingshot interactions 
with stars normally ensure that a binary system gets quickly 
closer, reaching separations $d_c \sim 1$ pc in less than $10^8$ 
yr \cite{beg80}. As the number of field stars with ideal impact 
parameters decreases with $d$, the binary evolution will then 
stall at around $d_c$ (i.e., well above the separation at which 
gravitational radiation becomes important), unless further 
angular momentum can be efficiently removed by other processes 
such as, for example, influx of gas or accretion disk interactions. 
SBBH systems are thus likely to spend most of their time at 
separations $d \sim (0.05-1)$ pc for BH masses of order $10^8\,
M_{\odot}$. While the origin of X-shaped radio morphologies in 
some radio galaxies may provide some phenomenological evidence for 
a possible spin flip during the final merger stage \cite{mer02},
thus suggesting that at least some SBBHs may coalesce within a 
Hubble time, it is theoretically still not yet fully understood 
today whether that can be achieved by a substantial fraction of 
SBBHs, e.g., see \cite{beg80,qui97,gou00,yu02,cha03,arm05,mer05}.
As shown below, our analysis presented here provides further 
phenomenological support for relatively short merging timescales 
in AGNs.

\section{Periodic variability in close SBBHs}
The presence of close SBBH systems has been repeatedly invoked
as plausible source for a number of observational findings in
blazar-type AGNs, ranging from misalignment and precession of 
jets to helical trajectories and quasi-periodic variability, 
see \cite{rie05a,kom06} for recent reviews. As illustration 
of the latter, consider a simple SBBH model with a precessing 
jet:\\
Interactions of the companion with the accretion disk around 
the primary BH may then naturally account for longterm optical 
periodicity with periods of the order of $P_{\rm obs}^{\rm opt} 
\sim $ several years as observed in a number of blazars, e.g., 
see \cite{val00} for OJ~287 and also Table~(\ref{tab1}), at 
least in those cases where the disk provides a non-vanishing 
contribution to the observed optical spectral flux. Accordingly, 
we may derive an upper limit for the intrinsic Keplerian orbital 
period of the binary
\beq 
 P_k \leq  \frac{2}{(1+z)}\,P_{\rm obs}^{\rm opt}\,
\eeq by assuming that the optical longterm periodicity is 
caused by the secondary BH crossing the disk around the 
primary twice per orbital period. Note, however, that due 
to internal disc warping and/or disc precession, some 
deviations from strict periodicity are likely to occur.\\
The detection of helical jet paths in an increasing 
number of blazar sources \cite{zen97,ran98,kel04,rie04} 
suggests that quasi-periodic variability -- especially in 
those energy bands dominated by the jet, e.g., radio, X- 
and $\gamma$-ray -- may also naturally arise as a consequence 
of differential Doppler boosting $S_{\nu}(t) = \delta(t)^{n} 
S_{\nu}'$ for a periodically changing viewing angle, where
$\delta(t)$ is the time-dependent Doppler factor, $n \geq 3$,
and $S_{\nu}$' the spectral flux in the comoving frame  
\cite{rie00,dep02}. For non-ballistic helical motion, classical 
travel time effects will then lead to a shortening of observable 
periods $P_{\rm obs}$ with respect to the real physical driving 
period $P$ such that 
\beq
 P_{\rm obs} \simeq (1+z)\,\frac{P}{\gamma_b^2}\,
\eeq where $\gamma_b \simeq (5-15)$ is the typical bulk flow 
Lorentz factor \cite{rie04}. Orbital motion and (Newtonian) 
disk precession caused by tidally induced perturbations in 
the disk \cite{kat97,rom00} belong to the most obvious driving 
sources for helical jet paths. If, as it is usually believed, 
the high energy emission is produced on small jet scales, it 
may be primarily modulated by the orbital SBBH motion, 
suggesting observable periods 
\beq\label{orbital}
 P_{\rm obs} \sim 30\,\left(\frac{P_{\rm obs}^{\rm opt}}
   {10\,{\rm yr}}\right)\,\left(\frac{15}{\gamma_b}\right)^2 
   \;\mathrm{d}\,.
\eeq as indeed observed [cf. Table~(\ref{tab1})]. On the other 
hand, simple cooling arguments suggest that a significant 
part of the radio emission may originate on larger scales, 
where Newtonian jet precession can no longer be neglected. 
In general, the driving period for Newtonian jet precession 
is (at least) an order of magnitude higher than the orbital 
period $P_k$, i.e., $P=P_p \geq 10 \,P_k$ \cite{rie04}, so 
that observable radio periods are expected to satisfy 
\beq
   P_{\rm obs}^{\rm radio} 
   \gppr 20 P_{\rm obs}^{\rm opt}/\gamma_b^2\,. 
\eeq Note, that if $P_p$ is rather small (say $P_p\sim 10\,P_k$), 
moderate bulk Lorentz factors are still sufficient to account 
for $P_{\rm obs}^{\rm radio} < P_{\rm obs}^{\rm opt}$ as 
sometimes observed, cf. Table~(\ref{tab1}).\\ 
Let us note, that a close SBBH interpretation for the origin of
(some) quasi-periodic variability is certainly not the only 
possible explanation, as other (not mutually exclusive!) origins 
(e.g., disk instabilities, orbiting disk hot spots) are 
conceivable as well. 
What makes the SBBH interpretation unique, however, is that it 
seems to allow corroboration of facts otherwise not possible:
(1.) It is based on quite general arguments for (bottom-up) 
structure formation (galaxy mergers), (2.) it naturally accounts 
for helical jet trajectories observed in many sources, (3.) it 
can offer a reasonable solution to the problem of divergent 
central mass estimates derived from high energy emission models 
and host galaxy observations \cite{rie03}, and (4.) it provides 
a coherent explanation for longterm periodic variability. In 
particular, as shown above, strong support can be provided by 
the detection of quasi-periodic variability on different 
timescales in different energy bands. 
\begin{table*}
  \begin{center}
  \caption{Properties of a sample of blazar SBBH candidates,
           cf. Rieger et al., in preparation for more details,
           with masses in $10^8\, M_{\odot}$, and upper limits 
           for the associated binary lifetime $\tau_{\rm grav}$ 
           due to gravitational radiation in units of $10^8$ yr 
           for typical mass ratios $\geq 0.01$. References given
           are for observed periods $P_{\rm obs}$ identified.}
  \vspace*{0.0cm}
  \label{tab1}
  \begin{tabular}{lclccccc}\hline
    name     & ~redshift $z$~ &  periods $P_{\rm obs}$  &  Ref. &
             $(m+M)$ & $P_k$ [yr] & $d/10^{16}$cm ~&~$\tau_{\rm grav}$ \\\hline
    Mkn~501  & 0.034  & 23.6 d (X-ray)      &  \cite{kra99}   & (2-7)  &(6-14)     & (2.5-6)  & 5.50 \\
             &        & $\sim 23$ d (TeV)   &  \cite{oso06}   &        &           &          &      \\
    BL Lac~  & 0.069  & $13.97$ yr (optical)&  \cite{fan98b}  &(2-4)   &(13-26.1)  & (4.8-9.7)& 28.9 \\
             &        & $\sim 4$ yr (radio) & \cite{kel03}    &        &           &      & \\
    ON 231   & 0.102  & $\sim 13.6$ yr (optical)& \cite{liu95}& $\geq 1$& (12.3-24.7)& $\geq$ 3.7 & 79.4 \\  
             &        & $\sim 3.8$ yr (optical) & \cite{bel00} &        &          &      & \\
    3C~273~  & 0.158  & 13.65 yr (optical)      & \cite{fan01} & (6-10) &(11.8-23.6)& (6.5-12.3) & 3.55 \\
             &        & 8.55 yr (radio)         & \cite{cia04} &        &           &          &      \\
    OJ~287~  & 0.306  & 11.86 yr (optical)     & \cite{sil88} & 6.2   & (9.1-18.2) & (5.5-8.8)& 1.68 \\
             &        & $\sim 12$ yr (infrared)& \cite{fan98a}&       &           &          & \\
             &        & $\sim 1.66$ yr (radio) & \cite{hug98} &       &           &          & \\
             &        & $\sim 40$ d (optical)  & \cite{wu06}  &       &           &          & \\
    3C~66A   & 0.444  & 4.52 yr (optical)      & \cite{fan02} & $\geq 1$&(3.1-6.3) & $\geq$ 1.5 & 2.08\\
             &        & 65 d (optical)         & \cite{lai99} &         &          &          &      \\
    AO 0235  & 0.940  & 2.95 yr (optical)?     & \cite{fan02} &$\geq 1$ &(1.5-3.1) &$\geq$ 0.95 & 0.31\\
             &        & 8.2 (optical)?         & \cite{rai06} &      &  (4.2-8.4)  &$\geq$ 1.81 & 4.48\\
             &        & 5.7 yr (radio)         & \cite{rai01} &      &             &           &\\
    3C~446   & 1.404  & 4.7 yr (optical)       & \cite{web88} &  6.5 &  (1.9-3.9)  & (2.0-3.2) & 0.03\\
             &        & 5.8 yr (radio)         & \cite{kud06} &      &             &           & \\\hline
 \end{tabular}
 \end{center}
 \end{table*}
\section{Application to individual sources}
\subsection{AO 0235+16:} 
Long-term monitoring (1975-2000) of this well-known and highly 
variable BL Lac object has shown evidence for a $(5.7\pm 0.5)$ 
yr radio and a possible $(2.95 \pm 0.15)$ or $(5.7\pm 0.5)$ yr 
optical periodicity \cite{rai01,fan02}. Both findings have been 
interpreted within a close SBBH framework \cite{rom03,ost04}, 
cf. \cite{rie05b} for a discussion. However, during the latest 
radio to optical monitoring campaign in 2003-2005, no evidence 
for a major radio or optical outburst -- extrapolated from 
previous observational results to occur within the campaign -- 
was found \cite{rai05,rai06}. 
While the non-detection of a major radio outburst might be 
relatively easy accommodated in a precession-driven helical 
jet model \cite{rom03} by taking, for example, a moderate change 
of the jet bulk Lorentz or Doppler factor (decrease) and/or the 
inner disk properties during an active stage fully into account, 
the non-detection of the expected optical outburst (if indeed 
associated with accretion disk interactions, e.g., \cite{rom03}),
is somewhat more challenging for a consistent SBBH interpretation. 
A possible way out -- apart from the possibility of inaccurate 
previous periodicity results, i.e., the real optical period may 
actually be $\sim 8$ yr \cite{rai06}, which needs to be checked 
in detail -- is to assume that, similar to the accretion-ejection 
connection in microquasars, an active (low-hard-type) source stage 
may be associated with enhanced intrinsic (!) jet activity and a 
decrease in optical disk flux, so that the corresponding disk 
contribution might be swamped by emission from the jet. If this 
is the case, then one might expect some orbital-driven optical 
variability [cf. eq.~(\ref{orbital})] to occur on timescales 
$\gppr 20$ d assuming $\gamma_{b,\,{\rm optical}} \simeq 8$ for 
the optical regime \cite{zha02}. Clearly, extensive 
multiwavelength monitoring of this source around the next 
SBBH anticipated outburst and advanced analysis methods will
be important to clarify whether the latest anomaly may turn 
into a falsification of a simple SBBH scenario, where optical 
longterm periodicity is related to disk crossing. 
We note that circumstantial evidence for very high Doppler 
factors $\sim 100$ ($\gamma_b \geq 50$) in AO 0235 have been 
reported in the literature, e.g., see \cite{fre06}, which -- 
if indeed true -- would even allow for some orbital-driven 
intraday variability.
A very interesting alternative explanation for the origin of 
some radio QPOs in AO 0235+16 (including its $\sim 5.7$ yr 
periodicity) has been proposed recently \cite{liu06}, suggesting 
that they might be related to periodic plasma injection into 
the jet driven by the p-mode oscillation of a thick inner disk 
(probably excited by a close SBBH) with a relatively high 
transition radius $R_{\rm tr} \sim 10^3 r_g$ (cf. however also 
the general arguments for much smaller transition radii in BL 
Lacs, \cite{cao03}) and associated intrinsic fundamental 
frequency $f_0 \simeq 0.18 \,\mathrm{yr}^{-1} (R_{\rm tr}/1.2 
\cdot 10^3 r_g)^{-3/2}\,(4.7\cdot 10^8\,M_{\odot}/M_{\rm BH})$. 
As the QPOs are thought to arise due to periodic plasma injection, 
the model seems to imply that similar QPOs should be observable 
in different jet-dominated energy bands (i.e, not only in the 
radio), which may allow a straightforward test. To complicate 
matters, note however that due to the above noted travel-time 
effects the observed radio periods may not necessarily 
correspond to the intrinsic driving periods.
\subsection{OJ~287:}
Optical and infrared monitoring of this famous BL Lac object 
have shown strong evidence for a $\sim 12\,$ yr 
\cite{sil88,fan98b} longterm, and a possible $\sim 40$ d 
midterm periodicity \cite{wu06}, with the 12 yr periodicity 
commonly interpreted as due to a close SBBH system, e.g., 
see \cite{sil88,val00,liu02}. Assuming the 12~yr QPO to be 
caused by disk crossing implies $P_k \lppr 18.2$ yr. 
Interestingly, orbital-driven helical jet motion 
[Eq.~(\ref{orbital})] then suggests observable midterm 
periodicity $P_{\rm obs} \lppr 50$ d for $\gamma_b \sim 13$ 
as derived from SED multiwavelength modelling of OJ~287 
\cite{pad01}, which is well consistent with the observed 
timescale. If the radio emission emerges from larger 
Newtonian-precession-modulated scales with $P_p \sim 10\,
P_k$, radio QPOs with $P_{\rm obs} \sim 1.5$ yr might be 
expected assuming a similar bulk Lorentz factor for the 
radio regime. Again, there is significant evidence for such 
a periodicity in the radio data \cite{hug98}, suggesting 
that a close SBBH may indeed offer a powerful explanatory 
framework.

\section{Jet activity and evolution of SBBHs}
Table~(\ref{tab1}) presents properties derived for a sample 
of blazar SBBH candidates, where the observed periodicities have 
been used to estimate the last three columns: In all cases the 
gravitational lifetime of the binary is (a) significantly smaller 
than the Hubble time for the most likely range of mass ratios, 
a result that gives strong phenomenological support to the notion 
that SBBHs can indeed coalesce, and (b) still large enough to 
satisfy the minimum source lifetime required for jet fuelling 
(e.g., $\tau \sim 10^6$ yr for the quasar 3C~273). The derived 
separations $d$ are of order of the maximum size $r_d \sim 1000
\,r_g$ (as given by the Toomre stability condition $Q \simeq 1$) 
for a standard Shakura \& Sunyaev disk around the 
primary.\footnote{Note that higher values for $r_d$, perhaps 
even up to one parsec, may be possible if magnetic torques from 
disk winds are fully taken into account, cf. \cite{goo03}.} 
Accordingly, our results suggest that the candidate sources may 
undergo early stages of binary-disk interaction, where the orbit 
of the secondary BH is still inclined with respect to the 
circumprimary disc \cite{iva99}. This may qualify our assumption 
that the secondary BH indeed hits the disk around the primary 
twice per orbital period, so that the larger (upper bound) values 
for $P_k$ and $d$ in Table~(\ref{tab1}) appear to be more realistic.
Accretion disk interactions will lead to an accelerated evolution 
of the binary and can also excite the binary to eccentricities $e 
\sim 0.1$ \cite{iva99,arm05}. 
Gas accretion and binary-disc interactions are thus very likely 
to dominate the most critical binary evolution stage between $d 
\sim 0.01-1$ pc, supporting the notion that SBBH systems are 
important but temporary feature of galaxy evolution, e.g., 
\cite{gou00}. Binary-disc interactions can also provide a natural 
trigger for enhanced jet activity (e.g., time-dependent increase of 
accretion rate, cf. \cite{sil88}) and may even lead to recurrent 
ejection of superluminal jet components. Moreover, during collision 
the disk gas will be perturbed, shocked, heated and ejected at the 
point of collision, which may lead to the possible formation of hot, 
optically thick outflows from the disk with maximum luminosity up 
to the Eddington limit of the secondary \cite{iva98}.

\section{Conclusion}
While each piece of observational evidence (e.g., helical jet 
trajectories, periodic variability etc.) may in principle allow 
a variety of alternative interpretations, we believe (cf. $\S 2$) 
that a strong cumulative (phenomenology-based) case can be built 
for the presence of close supermassive binary black holes at the 
centers of (at least some) radioloud Active Galaxies. If so, then
simple AGN unification schemes may be fundamentally incomplete as, 
for example, activity cycles, jet structure and emission properties 
may be strongly affected by the presence of a secondary BH. 
Longterm monitoring of blazar sources, the use of astronomical 
plate archives and sophisticated analysis methods may thus be 
crucial for further in-depth assessment of the robustness of the 
observational basis.

\begin{acknowledgements}
Partial support by a Cosmogrid Fellowship and useful comments by
the referee are gratefully acknowledged.
\end{acknowledgements}


\begin{thebibliography}{3}
\bibitem{beg80} Begelman, M.C., Blandford, R.D., Rees, M.J.: 
                Nature \textbf{287}, 307 (1980)
\bibitem{fer05} Ferrarese, L., Ford, H.: SSRv \textbf{116}, 523 (2005)
\bibitem{lob05} Lobanov, A.P.: Mem. S.A.It. \textbf{76}, 164 (2005)
\bibitem{dre97} Dressler, A. et al.: ApJ \textbf{490}, 577 (1997)
\bibitem{kom03} Komossa, S., et al.: ApJL \textbf{582}, L15 (2003)
\bibitem{rod06} Rodriguez, C., et al.: ApJ \textbf{646}, 49 (2006)
\bibitem{mer02} Merritt, D., Ekers, R.~D.: 2002, Science 
                \textbf{297}, 1310 (2002)
\bibitem{qui97} Quinlan, G.~D., Hernquist, L.: New Astronomy 
                \textbf{2}, 533 (1997) 
\bibitem{gou00} Gould, A., Rix, H.-W.: ApJL \textbf{532,} L29 (2000)
\bibitem{yu02} Yu, Q.: MNRAS \textbf{331}, 935 (2002)
\bibitem{cha03} Chatterjee, P., et al.: ApJ \textbf{592}, 32 (2003)
\bibitem{arm05} Armitage, P.J., Natarajan, P.: ApJ \textbf{634}, 921 (2005)
\bibitem{mer05} Merritt, D., Milosavljevi{\'c}, M.: Living Reviews in 
                Relativity \textbf{8}, 8 (2005)
\bibitem{rie05a} Rieger, F.M.: in Proc. 22nd Texas Symposium on Relativistic
                 Astrophysics (Stanford 2004), eds. P. Chen et al.
                 (eConf:C041213), 1601 (2005a)
\bibitem{kom06} Komossa, S.: Mem. S.A.It \textbf{77}, 733 (2006)
\bibitem{val00} Valtaoja, E., et al.: ApJ \textbf{531}, 744 (2000)
\bibitem{zen97} Zensus, A.: ARA\&A \textbf{35}, 607 (1997)
\bibitem{ran98} Rantakyro, F.T., et al.: A\&AS \textbf{131}, 451 (1998)
\bibitem{kel04} Kellermann, K.I., et al.: ApJ \textbf{609}, 539 (2004) 
\bibitem{rie04} Rieger, F.M.: ApJL \textbf{615}, L5 (2004)
\bibitem{rie00} Rieger, F.M., Mannheim, K.: A\&A \textbf{359}, 948 (2000)
\bibitem{dep02} De Paolis, F., et al.: A\&A \textbf{388}, 470 (2002)
\bibitem{kat97} Katz, J.: ApJ \textbf{478}, 527 (1997)
\bibitem{rom00} Romero, G., et al.: A\&A \textbf{360}, 57 (2000)
\bibitem{rie03} Rieger, F.M., Mannheim, K.: A\&A \textbf{397}, 121 (2003)
\bibitem{rai01} Raiteri, C.M., et al.: A\&A \textbf{377}, 396 (2001)
\bibitem{fan02} Fan, J.H., et al.: A\&A \textbf{381}, 1 (2002) 
\bibitem{rom03} Romero, G.E., Fan, J., \& Nuza, S.E.: ChJAA \textbf{3}, 
                513 (2003)
\bibitem{ost04} Ostorero, L., Villata, M., Raiteri, C.M.: A\&A \textbf{419}, 
                913 (2004)
\bibitem{rie05b} Rieger, F.M.: AIP Conf. Proc. \textbf{745}, 487 (2005b)
\bibitem{rai05} Raiteri, C.M., et al.: A\&A \textbf{438}, 39 (2005) 
\bibitem{rai06} Raiteri, C.M., et al.: A\&A \textbf{459}, 731 (2006) 
\bibitem{zha02} Zhang, L.Z., Fan, J.H., Cheng, K.S.: PASJ \textbf{54}, 
                159 (2002)
\bibitem{fre06} Frey, S., et al.: PASJ \textbf{58}, 217 (2006)
\bibitem{liu06} Liu, F.K., Zhao, G., Wu, X-B.: ApJ \textbf{650}, 749 (2006)
\bibitem{cao03} Cao, X.: ApJ \textbf{599}, 147 (2003)
\bibitem{sil88} Sillanp{\" a}{\" a}, A., et al.: ApJ \textbf{325} 628 (1988)
\bibitem{fan98b} Fan, J.H., et al.: ApJ \textbf{507}, 173 (1998b) 
\bibitem{wu06} Wu, J., et al.: AJ \textbf{132}, 1256 (2006) 
\bibitem{liu02} Liu, F.K., Wu, X-B.: A\&A \textbf{388}, L48 (2002)
\bibitem{pad01} Padovani, P., et al.: MNRAS \textbf{328}, 931 (2001)
\bibitem{hug98} Hughes, P.~A., Aller, H.~D., Aller, M.~F.: ApJ \textbf{503}, 
                662 (1998) 
\bibitem{kra99} Kranich, D., et al.: 26th ICRC (Salt Lake City) \textbf{3}, 
                358 (1999) 
\bibitem{oso06} Osone, S: Astroparticle Physics \textbf{26}, 209 (2006) 
\bibitem{kel03} Kelly, B.C., et al.: ApJ \textbf{591}, 695 (2003)
\bibitem{liu95} Liu, F.K., Xie, G.Z., Bai, J.M.: A\&A \textbf{295}, 1
\bibitem{bel00} Belokon, E.T., Babadzhanyants, M.K., Pollock, J.T.: A\&A 
                \textbf{356}, L21 (2000)
\bibitem{fan01} Fan, J.H., Romero, G.E., Lin, R.: ChA\&A \textbf{25}, 282 
                (2001) 
\bibitem{cia04} Ciaramella, A., et al.: A\&A \textbf{419}, 485 (2004) 
\bibitem{fan98a} Fan, J.H., et al.: A\&AS \textbf{133}, 163 (1998a)
\bibitem{lai99} Lainela, M., et al.: ApJ \textbf{521}, 561 (1999) 
\bibitem{web88} Webb, J.R., et al.: AJ \textbf{95}, 374 (1988) 
\bibitem{kud06} Kudryavtseva, N.A., Pyatunina, T.B.: Astronomy Reports 
                \textbf{50}, 1 (2006) 
\bibitem{goo03} Goodman, J.: MNRAS \textbf{339}, 937 (2003)
\bibitem{iva99} Ivanov, P.B., Papaloizou, J.C.B., Polnarev, A.G.: MNRAS 
                \textbf{307}, 79 (1999)
\bibitem{iva98} Ivanov, P.B., Igumenshchev, I.V., Novikov, I.D.: ApJ 
                 \textbf{507}, 131 (1998)
\end{thebibliography}
\end{document}